\documentstyle[12pt]{article}

\begin{document}
\hfill{UTTG-08-92}\\
\hspace*{\fill} {UR-1254}\\
\hspace*{\fill} {ER-40685-706}\\
\vspace{24pt}
\begin{center}
{\bf High Temperature Partition Function of the Rigid String}

\vspace{24pt}
Joseph Polchinski

\vspace{12pt}
Theory Group\\ Department of Physics \\ University of Texas
\\ Austin, Texas 78712\\ bitnet:joe@utaphy

\vspace{18pt}
Zhu Yang

\vspace{12pt}
Department of Physics and Astronomy\\ University of Rochester\\
Rochester, New York 14627\\ bitnet:yang@uorhep

\vspace{12pt}
{\bf ABSTRACT}

\end{center}
\baselineskip=21pt

\begin{minipage}{4.8in}
\baselineskip=21pt
We find that the high temperature limit of the free energy per unit
length for the rigid string agrees dimensionally with that of the QCD
string (unlike the Nambu-Goto string).  The sign, and in fact the
phase, do not agree.  While this may be a clue to a string theory of QCD,
we note that the problem of the fourth derivative action makes it
impossible for the rigid string to be a correct description.

\end{minipage}

\vfill

\pagebreak

\setcounter{page}{1}

It is an old idea that large-$N$ QCD might be exactly reformulated as
a string theory[1].  Recently, one of us has shown that the partition
function of a QCD flux tube has a natural continuation past the
deconfining transition, and that the high-temperature continuation can be
evaluated perturbatively[2]. Thus, we have at least one analytic fact about
the QCD flux tube, and any equivalent string theory must correctly
reproduce this.  In ref.~[2], the result was compared with the continued
partition function of the Nambu-Goto string, and it was found that the two
agree only if the string has an infinite number of massive world-sheet
degrees of freedom, the effective number growing as the square of the
temperature. This is very plausible, corresponding to internal
excitations of the transverse structure of the tube.

However, this is not the
only possible interpretation.  The calculation in ref.~[2] is necessarily
in an unphysical region, and so does not {\it directly} measure the number
of degrees of freedom.  A modification of the short-distance structure of
the string theory could also produce the same continuation.  Indeed, this
point has recently been made independently by Green[3], who shows that his
long-standing proposal to obtain partonic structure from Dirichlet
conditions on string boundaries may meet the test of ref.~[2].

A different modification is the rigid string theory
proposed in refs.~[4,5].  Addition of the higher-derivative extrinsic
curvature term leaves the world-sheet theory renormalizable but
significantly changes the short-distance properties.  The purpose of this
note is to examine the effect of this term on the high-temperature
continuation. Actually, as we will discuss briefly later, the extrinsic
curvature idea would seem to have a serious problem at a much more basic
level.  However, it provides a useful example for developing the method of
ref.~[2].

The conformal-gauge
action is[4]
\begin{equation}
S = \frac{1}{2\alpha_0} \int d^2 \sigma\, \Bigl\{ \rho^{-1} \partial^2 X^\mu
\partial^2 X_\mu
+ \lambda^{ab} (\partial_a X^\mu \partial_b X_\nu - \rho
\delta_{ab} ) \Bigr\}
+ \mu_0 \int d^2\sigma\, \rho.
\end{equation}
The field $\lambda_{ab}$ is a Lagrange multiplier, setting the induced
metric $\partial_a X^\mu
\partial_b X_\mu$ equal to the intrinsic metric $\rho
\delta_{ab}$.  The dimensionless coupling $\alpha_0$ is asymptotically
free[6]. The explicit string tension $\mu_0$ is now unimportant at high
energy but cuts off the running of $\alpha_0$ at low energy.  The
Liouville term from the Faddeev-Popov determinant can be omitted: it is
comparable to the {\it two-loop} effect of $\alpha_0$.\footnote{This is
true in the high-energy, curvature-dominated region[4].  In the low energy,
tension-dominated region studied in ref.~[7], the Liouville term is more
important than the extrinsic curvature term.}

The partition function of this theory near the deconfining transition
has been extensively studied by Kleinert and German[8].  This work is
complementary to ours: it is in a physical region, but a complicated one
where QCD must be treated by Monte-Carlo methods.  In contrast, we look in an
unphysical region where analytic calculation is possible.  Although our
calculation overlaps with that of ref.~[8], we find it useful to begin from
scratch, because it is necessary for the purpose of continuation to have an
analytic result at all temperatures, while the results of ref.~[8] are in
part numerical.

In order to make an analytic treatment possible, we will consider the case
that $\alpha_0$ is small at all scales.  We are interested in the partition
function in the presence of a source and anti-source separated by a large
distance $L$. The world-sheet is an annulus, with a modulus $t$ that we
will incorporate by choosing the coordinate $\sigma_1$ to run from 0 to
$L$, and the periodic coordinate $\sigma_2$ to run from 0 to $\beta t$.
We expand around the configuration
\begin{eqnarray}
X^1 (\sigma) = \sigma^1 && X^0 (\sigma) = \sigma^2/t \nonumber\\[2pt]
\rho(\sigma) = \bar\rho && \lambda^{11}(\sigma) = \bar\lambda^{11} \qquad
\lambda^{22}(\sigma) = \bar\lambda^{22}.
\end{eqnarray}
The effective action is $S_{\rm eff} = S_0 + S_1 + \ldots$, where the
tree-level action is
\begin{equation}
S_0 = \frac{ L \beta t}{2 \alpha_0} \Bigl\{ \bar\lambda^{11} +
\bar\lambda^{22} t^{-2} + \bar\rho (2 \alpha_0 \mu_0 - \bar\lambda^{aa})
\Bigr\}
\end{equation}
and the one-loop action is
\begin{eqnarray}
S_1 &=& \frac{d - 2}{2} \ln \det (\partial^4 - \bar\rho \bar\lambda^{ab}
\partial_a
\partial_b) \nonumber\\[2pt]
&=& \frac{d-2}{2} L \sum_{n = -\infty}^\infty \int_{-\infty}^{\infty}
\frac{dk_1}{2\pi}\, \ln\Bigl( [ k_1^2 + 4\pi^2 n^2 \beta^{-2} t^{-2}
]^2\nonumber \\
&&\qquad\qquad\qquad\qquad\qquad
+ \bar\rho [ \bar\lambda^{11} k_1^2 + 4\pi^2 n^2 \bar\lambda^{22}
\beta^{-2} t^{-2}]  \Bigr) \label{eq:xx}
\end{eqnarray}
Note that only the transverse $X^\mu$ make a net contribution to the
one-loop term.  The two-loop and higher terms are suppressed by powers of
$\alpha_0$.\footnote{This is not the same as the perturbative
expansion criticized in ref.~[9].  The latter treated the one-loop
term as small compared to the tree-level term, which is not correct at high
temperatures $\beta^{-2} >> \mu_0$.}

We must extremize the action with respect to $t$,
$\bar\rho$, and $\bar\lambda^{ab}$.  The one-loop term cannot be evaluated
in simple form, but for small $\alpha_0$ there are two simple
approximations of overlapping validity.  At low temperature, $\beta^{-2} <<
\mu_0$, the one-loop term is small compared to the tree-level term and the
extremum lies close to the classical point $t = \bar\rho = 1$,
$\bar\lambda^{11} = \bar\lambda^{22} = \alpha_0\mu_0$.  At high
temperature, $\beta^{-2} >> \alpha_0 \mu_0$, the second derivative
term in the logarithm~(\ref{eq:xx}) is small compared to the fourth
derivative term, and we can expand
\begin{equation}
S_1 = \frac{d-2}{2} L \Biggr\{ -\frac{2\pi}{3\beta t} + (\bar\rho
\bar\lambda^{11})^{1/2} + O(\beta) \Biggr\}. \label{eq:1l}
\end{equation}
The first term in braces is the effective high-temperature action for
$2(d-2)$ massless scalars, the doubling coming from the
fourth-derivative kinetic term.  The second term comes only from $n = 0$ in
the frequency sum.

One can now extremize $S_0$ + $S_1$ to find
\begin{eqnarray}
\bar\rho^{-1} &=& t^2 \ =\ 1 - \frac{( d-2 ) \alpha_0}{2\beta}
(\bar\lambda^{11})^{-1/2} \nonumber\\[2pt]
\bar\lambda^{22} &=& 2\alpha_0\mu_0 - \bar\lambda^{11} + \frac{( d-2 )
\alpha_0}{2\beta} (\bar\lambda^{11})^{1/2}
\end{eqnarray}
with $\bar\lambda^{11}$ determined by
\begin{equation}
\bar\lambda^{11} - \frac{3(d-2)\alpha_0}{4\beta} (\bar\lambda^{11})^{1/2}
- \alpha_0 \mu_0 + \frac{\pi (d-2) \alpha_0}{3\beta^2} = 0. \label{eq:quad}
\end{equation}
The winding-soliton mass $M_1(\beta) = \beta \mu(\beta) = S_{\rm eff}/L$,
is to this order
$ M_{1, \rm rigid}^2(\beta) =
{\beta^2 t^2 (\bar\lambda^{11})^2} / {\alpha_0^2} $.

At low temperature, $\beta^{-2} << \mu_0$, $M_{1, \rm rigid}
(\beta)$ is near its
classical value $\beta \mu_0$.  As the temperature increases, the values of
$\bar\lambda^{11}$ and $t^2$ decrease until at
\begin{equation}
\beta_c^2 \mu_0 =
\pi(d - 2)/3 - (d-2)^2 \alpha_0/8
\end{equation}
the value of $t^2$ and thus of $M_{1, \rm rigid}^2$
goes through zero.  The winding state becomes tachyonic, signalling the
Hagedorn transition.\footnote{To obtain the full $O(\alpha_0)$ correction
to $\beta_c$, for comparison with figure~4 of ref.~[8], it is necessary to
keep the next correction to the high-temperature result~(\ref{eq:1l}). In
effect this correction amounts to a small renormalization of $\mu_0$ but
does not alter the qualitative behavior of the solution.}
In the Nambu-Goto and QCD strings, this is the end of the story, but the
situation here is more complicated.  At the slightly higher temperature
\begin{equation}
\beta_{c'}^2 \mu_0 = \pi(d - 2)/3 - 9 (d-2)^2 \alpha_0/64,
\end{equation}
the roots of
the quadratic~(\ref{eq:quad}) become complex, and at high temperatures
$\bar\lambda^{11}$ becomes large and predominantly imaginary.  The high
temperature limit of the soliton mass-squared is
\begin{equation}
M_{1, \rm rigid}^2(\beta) \sim \frac{\pi^2 (d-2)^2}{9\beta^2} \Bigl( 1 \pm i
[ 3\alpha_0 ( d-2 )/\pi]^{1/2} + O(\alpha_0, \mu_0\beta^2) \Bigr).
\label{eq:mo}
\end{equation}
This is to be compared with the large-$N$ QCD result[2]
\begin{equation}
M_{1, \rm QCD}^2(\beta) \sim -\frac{2g^2(\beta) N}{\pi^2\beta^2}.
\end{equation}
The Nambu-Goto result, in contrast, is $M_{1, \rm Nambu}^2(\beta) \sim
-(d-2)\pi\mu_0/3$.

We see that, unlike the Nambu-Goto string, the rigid string gives the
correct power of the temperature.  This may be because the UV
behavior of both theories is governed by a dimensionless coupling.  Note,
however, that we have found no direct connection between the
asymptotically free QCD and rigid string couplings: the QCD result
is proportional to $g^2(\beta) N$, which runs, while the rigid string
coefficient is a pure number.  More seriously, the two disagree in {\it
sign}.  The sign is curious---$M_{1\rm r}^2$ passes through zero at
the Hagedorn transition, but there is a second transition in the
rigid string.  Equally curious is the subleading imaginary part of
$M_{1, \rm rigid}^2$.  An imaginary part will appear due to the Hagedorn
instability, but at string {\it loop}, not tree, order.  The imaginary part
of $M_{1, \rm rigid}^2$ is due to a world-sheet, not spacetime, instability:
if we think of $\sigma_1$ as time, the field $X^\mu$
becomes unstable when $\bar\lambda^{11}$ changes sign.

World-sheet instabilities in the rigid string have previously been noted in
ref.~[10].
This brings up the serious problem of the consistency of the rigid string
theory.  It is well known that a fourth
derivative kinetic term must be
quantized either with an indefinite norm (this is the world-sheet
norm, but becomes the spacetime norm in the one-particle
sector) or with energy unbounded below.  Except for the
instability found in ref.~[10], this problem seems to have been ignored
in the literature.  Indeed, we would go even farther than ref.~[10]
in asserting that there can be {\it no} scale at which the dynamics can be
dominated by the curvature term.\footnote
{Therefore the asymptotic freedom of the rigid string
cannot be relevant to QCD,
since it occurs
only when the curvature term dominates.
This does not apply to the
statistical mechanics of random surfaces, where Minkowski continuation is
not an issue.  Also, it does not preclude a subleading
curvature term in a region
where the tension dominates.}  The point is that the spectral decomposition
of the $XX$ propagator on the world-sheet implies that $\partial \ln G
/ \partial \ln p \geq -2$ at all Euclidean momenta, but it would be roughly
$-4$ at a scale where the curvature term dominates.
Further, the negative norm states
certainly do not decouple here: they contribute to the partition function,
even in the physical region below the Hagedorn transition.

The main lesson of this work is that the QCD partition function
found in ref.~[2] might be compatible with a string theory with a finite
number of fields but with modified short-distance dynamics for the
$X^\mu$ fields.  However, the rigid string is also a reminder of how hard
it is to modify string theory in a consistent way.
Any consistent modification
is therefore of interest, and the idea of Green[3] deserves close
scrutiny.

\centerline{\bf Acknowledgements}

We would like to thank M. Green and P. Nelson
for discussions.
This research was supported in part by the
Robert A. Welch Foundation, NSF Grant PHY 9009850,  the Texas Advanced
Research Program, and U.S. Department of Energy Contract
No. DE-AC02-76ER13065.

\vfill

\pagebreak

\centerline{\bf References}

\begin{itemize}

\item[1.] G. 't Hooft, Nucl. Phys. {\bf B72}, 461 (1974).

\item[2.] J. Polchinski, Phys. Rev. Lett. {\bf 68}, 1267 (1992).

\item[3.] M. B. Green, {\it Temperature Dependence of String Theory in
the Presence of World-Sheet Boundaries}, Queen Mary College preprint
QMW-91-24 (1991).

\item[4.] A. M. Polyakov, Nucl. Phys. {\bf B268}, 406 (1986).

\item[5.] H. Kleinert, Phys. Lett. {\bf B174}, 335 (1986).

\item[6.] L. Peliti and S. Leibler, Phys. Rev. Lett. {\bf 54},1690
(1985).

\item[7.] J. Polchinski and A. Strominger, Phys. Rev. Lett. {\bf 67},
1681 (1991).

\item[8.] G. German, Mod. Phys. Lett. {\bf A6}, 1815 (1991).

\item[9.] G. German and H. Kleinert, Phys. Lett. {\bf B225}, 107 (1989).

\item[10.] E. Braaten and C. K. Zachos, Phys. Rev. {\bf D34}, 1512 (1987).

\end{itemize}

\end{document}